\documentclass[twocolumn,twoside,slac_two]{revtex4}
\usepackage{graphicx}
\usepackage{fancyhdr}
\begin{document}

\title{TeV Gamma-ray Blazar Monitoring Campaign}


\author{Eddie Collins-Hughes on behalf of the VERITAS collaboration}
\affiliation{University College Dublin, Belfield, Dublin 4}

\begin{abstract}
\noindent Presented here are the results from a search for flux variability in the data obtained on sources monitored with the Whipple 10m telescope from October 2010 to April 2011. The results of a search for correlation between detected emission from Markarian 421 and Markarian 501 with the Whipple 10m Telescope and Fermi-LAT detector will also be reported.
\end{abstract}

\maketitle

\thispagestyle{fancy}


\section{Introduction}

\noindent The Whipple 10 m telescope is an imaging atmospheric Cherenkov technique (IACT) instrument, located in southern Arizona, that operates in the 300 GeV to 10 TeV energy range. This instrument has been in operation since 1968 and was the first ground-based $\gamma$-ray telescope to detect a galactic source, the Crab Nebula in 1989 \cite{1}, and an extragalactic source, Markarian 421 in 1992 \cite{2}. Currently the telescope is operated by the VERITAS (Very Energetic Radiation Imaging Telescope Array System) collaboration as a dedicated blazar monitoring instrument, and is used as a trigger for VERITAS in case of enhanced activity from any of the blazars being monitored.\\

\noindent The Whipple 10 m telescope blazar monitoring campaign has been in operation since 1996, the scientific goals of the campaign are to examine the long and short-term variability of the emission from a selection of blazars.\\

\section{Blazars}

\noindent An Active Galactic Nucleus (AGN) is a galaxy for which the emission from its galactic core dwarfs that of the normal stellar component of the host galaxy. The galactic cores are believed to be powered by supermassive black holes surrounded by an accretion disk powering two jets of particles and electromagnetic radiation emitted perpendicular to the disk. Blazars are a sub-class of AGNs whose jet is at an observing angle of less than 10 degrees, making it the most obvious feature of the galaxy. Blazars exhibit broadband non-thermal variable emission with the presence of two peaks, one in the optical-keV range the other in the MeV-TeV range.\\

\noindent Emission models can generally be divided into two groups, where the particle species responsible for the $\gamma$-ray emission are either predominately leptonic or hadronic. Both model
families attribute the low-energy peak to synchrotron radiation from relativistic electrons within the jets. They differ on the origin of the MeV-TeV peak: leptonic models advocate the inverse-Compton scattering mechanism, utilising synchrotron-self-Compton (SSC) interactions or Compton interactions with an external photon field (e.g.\ see \cite{3}, \cite{4}, \cite{5}). Hadronic models account for the high-energy peak by neutral-pion or charged-pion decay with subsequent synchrotron and/or Compton emission from decay products, or synchrotron radiation from ultra relativistic protons (e.g.\ see \cite{6}, \cite{7})\\

\section{Scientific Motivation}
\noindent Long-term monitoring is needed to record both the long-term and short-term variability in the emission from blazars. Periods of intense, short-term variability are known as flares. These flares occur so rarely that a large amount of observing time must be dedicated to a handful of sources in order to increase the likelihood of seeing flaring activity. \\

%See Figure 3 for an example of the long term coverage possible with dedicated blazar monitoring campaigns.\\

\noindent Short-term variability is crucial to the calculation of the size of the photon emission region. The rise/fall time during a significant change in flux with temporal width 
\begin{math}
\Delta T
\end{math} 
provides an upper limit on the size of the emitting source given by 
\begin{math}
R < c \, \Delta T $/$ \delta
\end{math}, where 
\begin{math} R
\end{math}
 is the size of the emitting region and 
 \begin{math} c
\end{math}
  is the speed of light. \begin{math} \delta \end{math} is the Doppler factor, generally on the order of 10-30. Long-term monitoring is also needed in order to trigger multiwavelength (MWL) campaigns. When a source enters a flaring period VERITAS is alerted, along with a numerous MWL partners, to a target of opportunity (ToO). Broadband observations of correlated variability are critical for testing the predictions of different theoretical models.\\

\section{Results}

\noindent From April 2010 to October 2011 5 sources were monitored, with a total of 218 hours of data
collected. See Table 1 for each sources positional information, number of hours spent on source and significance level achieved.\\

\noindent The observations reported here were taken at an energy $>$ 400 GeV, the integral Crab flux above 400 GeV is (8.412+/-1.804\begin{math}_{stat}) \times 10^{-11} cm^{-2} s^{-1}\end{math}  \cite{8}. Lightcurves presented in this work are displayed in fluxes. These fluxes are calculated by normalising individual source rates to Crab units, where one Crab is equivalent to 1.25 \begin{math} \gamma \end{math}/min for the 10m telescope.\\

\noindent These data will form part of  a multi-year data set on Markarian 421 and Markarian 501. The Whipple 10m telescope's extensive coverage of these two sources at TeV energies is unique.\\

\begin{table}[ht!]
\begin{center}
\caption{Tabulation of data taken from October 2010 to April 2011. Hours is the number of hours taken on source, sigma is the significance level achieved.}
\begin{tabular}{|l|c|c|c|c|c|}
\hline \textbf{Source} & \textbf{RA} & \textbf{DEC} &\textbf{z} &\textbf{Hours}&\textbf{Sigma}\\
\hline 1ES 0229+200 &  02 32 39 & 20 17 17 & 0.14 & 41 & 0.8 \\
\hline Markarian 421 & 11 04 27 & 28 12 32 & 0.031 & 90 & 29.6 \\
\hline Markarian 501 & 16 53 22 & 39 45 36 & 0.034 & 33 & 19.2 \\
\hline 1ES 1959+650 & 19 59 59 & 65 08 55 & 0.047 & 12 & 4.1 \\
\hline 1ES 2344+514 & 23 47 04 & 51 42 18 & 0.044 & 42 & 1.9 \\
\hline
\end{tabular}
\label{l2ea4-t1}
\end{center}
\end{table}

\subsection{Markarian 421}

\noindent Markarian 421 is one of the most active VHE blazars and was the first to be
discovered at TeV energies. Its SED has peaks at keV and TeV energies and it has been known to demonstrate rapid, sub-hour scale flaring behaviour (e.g.\ see \cite{9}). It has a redshift of 0.031 making it the closest known TeV blazar. Fig. 1 shows the daily averaged lightcurve of Markarian 421 for 2010-2011, revealing clear day-scale variation (a $\chi^2$ test for constant emission returns a probability of $\sim 4.5 \times 10^{-31}$). A search within each night revealed no significant evidence for hour-scale variability.\\

\begin{figure}[ht!]
\centering
\includegraphics[width=85mm]{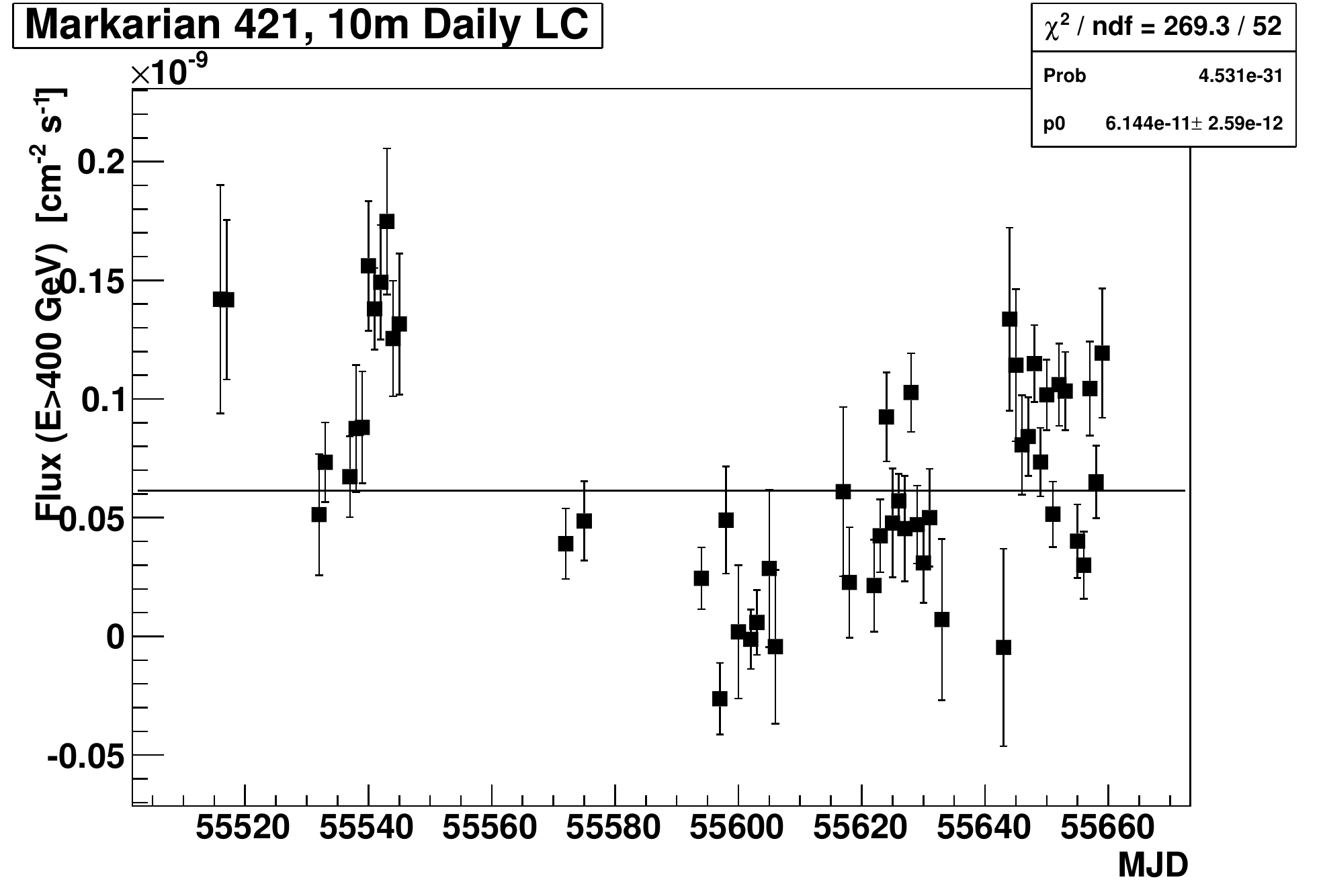}
\caption{The daily averaged lightcurve for Markarian 421 from the Whipple 10m telescope.}
\label{JACpic2-f1}
\end{figure}

\subsection{Markarian 501}

\noindent Markarian 501 also has a well cataloged history of flaring (e.g.\ see \cite{10}). Whilst being at a similar distance to Markarian 421, Markarian 501 is significantly weaker on average. This source has been intensively monitored in the past and it displays quite different temporal characteristics to Markarian 421. Fig. 2 shows the daily averaged lightcurve of Markarian 501 for 2010-2010. A $\chi^2$ test provides
evidence for significant day-scale variability (the probablility for the constant emission is
$\sim 5.1 \times 10^{-23}$). A search within each night revealed no significant evidence for hour-scale variability.\\

\begin{figure}[ht!]
\centering
\includegraphics[width=85mm]{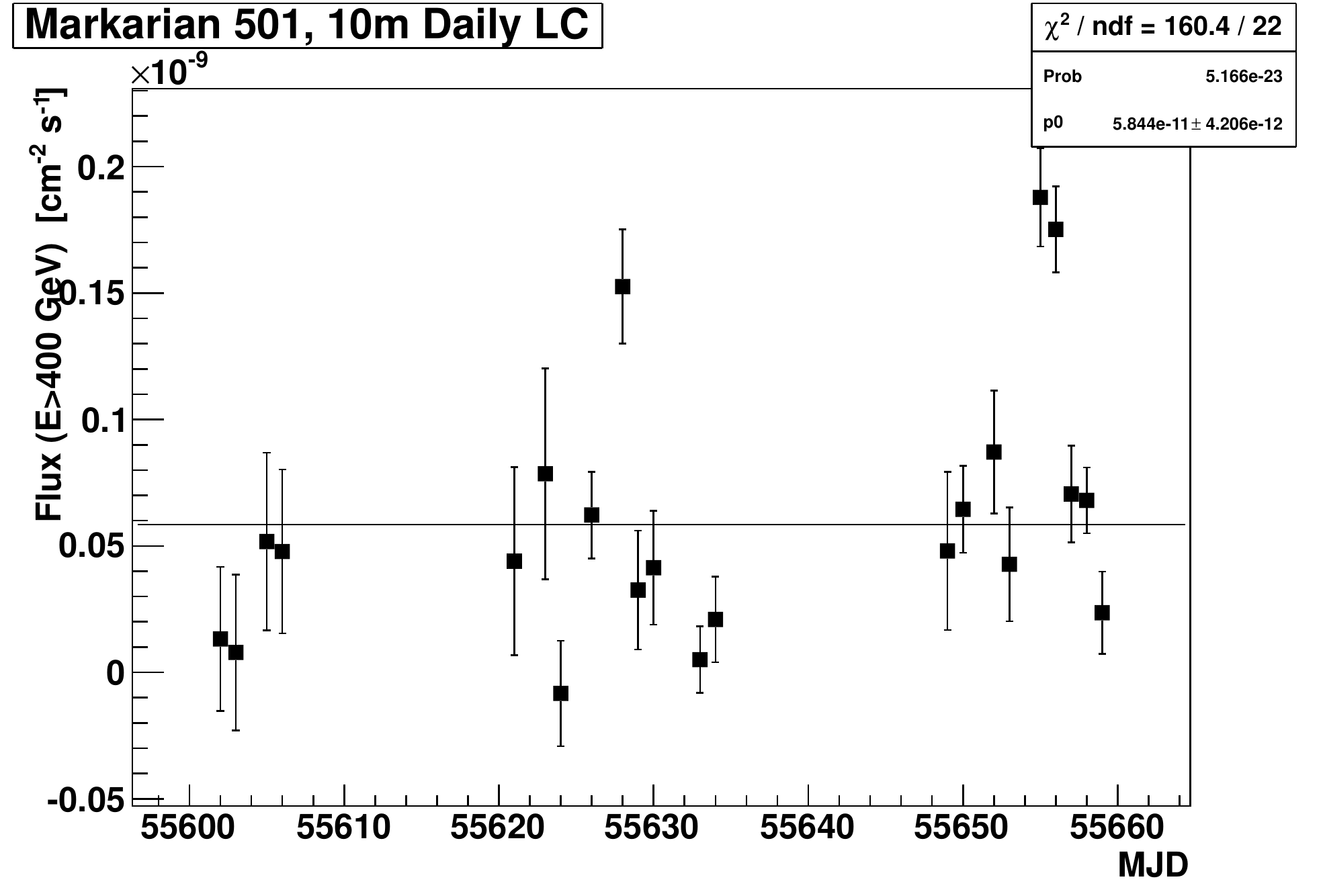}
\caption{The daily averaged lightcurve for Markarian 501 from the Whipple 10m telescope.}
\label{JACpic2-f1}
\end{figure}

\subsection{Other Sources}

\noindent1ES 1959+650 This source was discovered as a TeV gamma-ray emitter in
1998 by the Seven Telescope Array \cite{11}. In 2002, the Whipple 10m Telescope detected flaring activity from 1ES 1959+650 up to 5 times the Crab Nebula flux \cite{12}.\\

\noindent1ES 2344+514 The Whipple Collaboration reported a weak signal from the
object between 1999-2001 \cite{13}. As such it was chosen as a candidate likely to be observed again if it entered a period of enhanced activity.\\

\noindent1ES 0229+200 This blazar has an unusually hard TeV spectrum and as such
it is an excellent candidate for extragalactic background light (EBL) studies \cite{14}.\\

\noindent Neither 1ES 1959+650, 1ES 2344+514 nor 1ES 0229+200 have been detected with a significance level greater than 5 sigma during the 2010-2011 observing season and thus no statement about their variability can be made.\\

\section{Comparison with results from Fermi-LAT}

\noindent Shown in Fig. 3 are weekly averaged fluxes for Markarian 421 produced using the Whipple 10m and the publicly available Fermi-LAT data for the time period from October 2010 to April 2011.\\

\noindent
\begin{figure}[ht!]
\centering
\includegraphics[width=85mm]{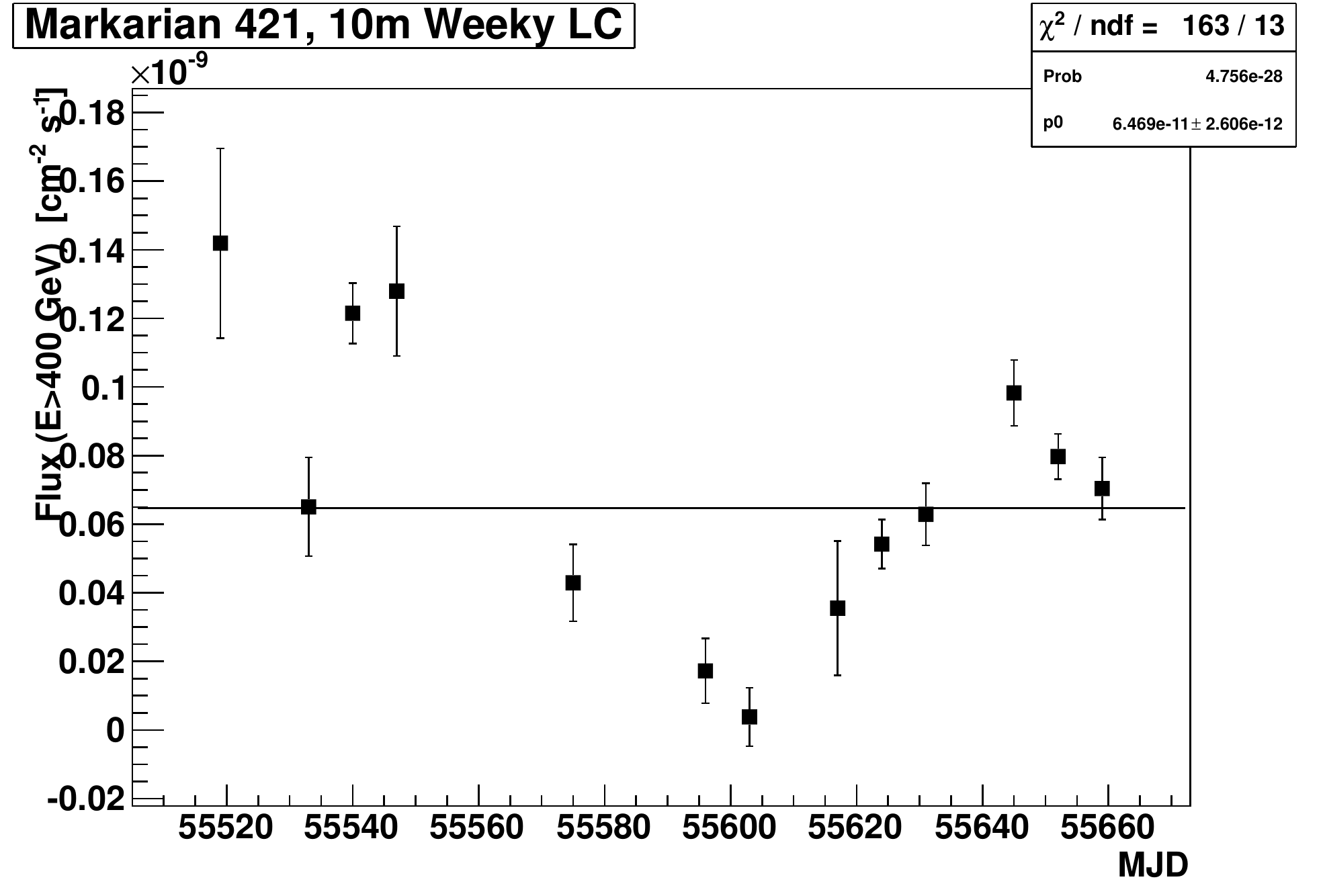}
\includegraphics[width=85mm]{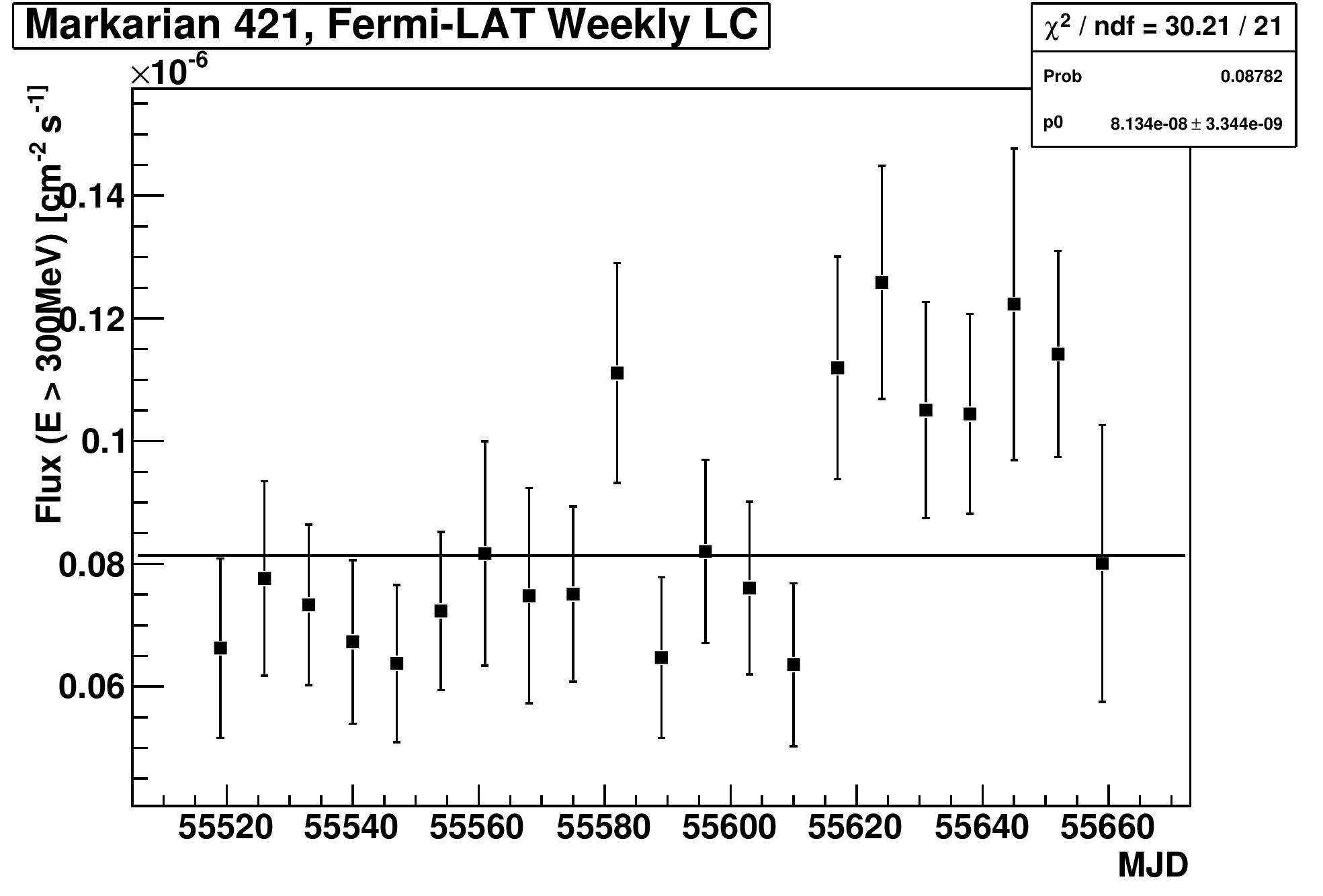}
\caption{Top graph is the weekly averaged fluxes from Markarian 421 as observed by the Whipple 10m telescope. Bottom graph is the weekly averaged fluxes from Markarian 421 as observed by Fermi-LAT. }
\end{figure}

\noindent Both the Whipple 10m telescope and Fermi-LAT see Markarian 421 as a variable emitter but no strong correlation in emission was found. A similar analysis performed for Markarian 501 also revealed no strong correlation. Discuss correlation test(extremely basic).\\

\section{Summary}

\noindent The Whipple 10m blazar monitoring campaign continued from October 2010 to June 2011, 210 hours of data were collected on five sources. Markarian 421 and 501 were both detected with high significance, and both demonstrated variability in their emission from night to night. Data from each night were examined for variability but no significant hour-scale variability was observed.\\

\noindent No strong correlation in emission from Markarian 421 or 501 as dectected by the Whipple 10 m telescope and the Fermi-LAT was observed.

\section{Acknowledgements}
\noindent This research is supported by grants from the U.S.
Department of Energy, the U.S. National Science Foundation, and the Smithsonian Institution, by NSERC in Canada, by STFC in the UK and by Science Foundation Ireland (SFI 10/RFP/AST2748).Research with the Whipple 10m gamma-ray telescope is supported by Fermi G.I. grant NNX10A048G\\

\end{document}